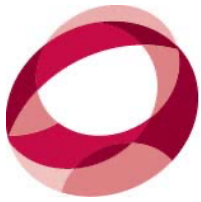

# Imagine All the People: Citizen Science, Artificial Intelligence, and Computational Research

*A Computing Community Consortium (CCC) Quadrennial Paper*

*Lea A. Shanley (University of Wisconsin-Madison; International Computer Science Institute, Berkeley, CA), Lucy Fortson (University of Minnesota), Tanya Berger-Wolf (The Ohio State University), Kevin Crowston (Syracuse University), Pietro Michelucci (Human Computation Institute)*

Machine learning, artificial intelligence, and deep learning have advanced significantly over the past decade. Nonetheless, humans possess unique abilities such as creativity, intuition, context and abstraction, analytic problem solving, and detecting unusual events. To successfully tackle pressing scientific and societal challenges, we need the complementary capabilities of both humans and machines. The Federal Government could accelerate its priorities on multiple fronts through judicious integration of citizen science and crowdsourcing with artificial intelligence (AI), Internet of Things (IoT), and cloud strategies.

***What are Citizen Science and Crowdsourcing?*** *Citizen science* describes a range of methodologies that support the meaningful contribution of members of the public to scientific and engineering research, as well as environmental monitoring. Although citizen science projects often engage the public in data collection or classification efforts, members of the public can contribute to all aspects of the scientific process. Despite the term, one does not need to be a citizen of any particular country to contribute to citizen science. Similarly, *crowdsourcing* engages a large group of people through an open call to tackle a common problem, either as individuals or collectively, sometimes incentivized through a prize or challenge. Related methodologies include "community science," "participatory science" "volunteer monitoring," and "volunteered geographic information." When designed and used appropriately, these approaches can augment traditional scientific methods and observing systems, for example, by providing or analyzing data at spatial and temporal resolutions or scales and speeds that otherwise would be impossible given limited staff and resources. **Projects are now exploring ways to incorporate AI techniques to complement public contributions, creating human-machine partnerships[1]** (see also Addendum)**.**

***Federal Government support for Citizen Science and Crowdsourcing.*** Federal agencies have embraced citizen science and crowdsourcing to advance their missions, improve delivery of government services, and promote a spirit of American volunteerism. While agencies like the National Science Foundation (NSF) and the National Institutes of Health (NIH) have long supported citizen science, a 2011 initiative launched by the Wilson Center's Commons Lab, with support from The Alfred P. Sloan Foundation, led to the establishment of the Federal Community of Practice for Crowdsourcing and Citizen Science (FedCCS), and co-development of the *Federal Citizen Science Catalog* with SciStarter, CitSci.org, and others. This initiative also sparked the *Crowdsourcing and Citizen Science Act* of 2016. From 2013 to 2014, the Commons Lab and FedCCS worked to garner high-level support for a step-by-step "How To" *Citizen Science Toolkit,* and a White House Office of Science and Technology Policy (OSTP) Memorandum in 2015 [2] [3]. The FedCCS, Catalog, and Toolkit served as the core components of *Citizenscience.gov*. Federal support has continued across Administrations through the growth of agency-specific communities of practice, adoption of new agency strategies and policies[4], and citizen science-focused funding opportunities.

---

[1] Ceccaroni, L., Bibby, J., et al. Opportunities and Risks for Citizen Science in the Age of AI. CSTP. (2019). 4(1), p.29. DOI: http://doi.org/10.5334/cstp.241 Available at:
https://theoryandpractice.citizenscienceassociation.org/articles/10.5334/cstp.241/

**Research Recommendations**

Federal support for integrating AI and citizen science will provide many opportunities for Americans to collectively Build Back Better, while rebuilding trust in science and government. We can build a post-pandemic workforce that leverages growing competency and comfort in collaborating with AI. We can build tools combining citizen science with AI and other technology to help communities and federal agencies collaboratively address challenges such as inequities in exposure to environmental pathogens, or to monitor and mitigate wildfires, flooding, drought, or other natural hazards. To seize these and other opportunities, we need to integrate Federal agency AI, Cloud, and citizen science strategies to address a research and development roadmap that gets us there.

**Human-Machine Teaming.** While capable of learning patterns beyond human capacity, AI systems can improve by learning from human analysis, referring uncertain cases to humans for help when needed. Human-machine teams can leverage human knowledge about the world to make sense of anomalies in the data that machines may miss. These abilities are particularly important in identifying potential biases inherent in the AI system that can further exacerbate systemic inequalities. Human perception and intuition also can quickly find solutions to some optimization problems that elude machines, which then serve as learning examples for improving the AI. Fundamental research is needed to learn how to best configure this human-machine teaming. Building on the CCC AI Roadmap[5] and the CCC Roadmap for Human Computation [6], core research challenges are:

- **Real-time data processing**: Citizen science is a proving ground for adaptive systems that leverage both humans and machines to process live sensor data in near real time. In particular, important gains can be made through research coupling citizen science use of AI on the Edge (installing pre-trained models on the sensors to filter data, recording only those instances pertinent to the citizen science task) and AI systems that can integrate heterogeneous data flows from citizen science deployed sensors via Open Data platforms that address privacy concerns.

- **Discovery and detecting rare events**: Detecting rare events (a form of "anomaly detection") can be cast as the problem of facilitating serendipitous discoveries in large data sets used for a task other than the one specified. Investigating the best combination of human and machine intelligences for anomaly detection can lead beyond scientific discoveries to improved algorithms that minimize the need for human intervention (e.g., in industrial settings).

---

[2] Shanley, L.A., Michelucci, P., Tsosie, K., Wyeth, G., Drapkin, J. K., Azelton, K., Cavalier, D., & Holmberg, J. (2021). Public Comment on Draft NOAA Citizen Science Strategy. *Human Computation*, 8, 25–42. https://doi.org/10.15346/hc.v8i1.130
[3] *White House Memorandum on Crowdsourcing and Citizen Science*. (2015) Available at https://obamawhitehouse.archives.gov/sites/default/files/microsites/ostp/holdren_citizen_science_memo_092915_0.pdf
[4] *NOAA Citizen Science Strategy*. (2021). Available at https://nrc.noaa.gov/Portals/0/Citizen%20Science%20Strategy%20_final.pdf?ver=2021-01-15-103436-693
[5] Gil, Y., and Selman, B., *A 20-Year Community Roadmap for Artificial Intelligence Research in the US,* Computing Community Consortium (CCC) and Association for the Advancement of Artificial Intelligence (AAAI), August 2019. arXiv:1908.02624 https://cra.org/ccc/resources/workshop-reports/
[6] Michelucci, P., Shanley, L.A., Dickenson, J., Hirsch, H. A, Bloomberg, M., Witbrock, M., *US Human Computation Roadmap*, Computing Community Consortium (CCC). (2014). https://cra.org/ccc/wp-content/uploads/sites/2/2015/05/Final-HC-Report.pdf

- **Improving data quality**: The better the quality and variety of data submitted by humans, the better the training data are for AI. Research is needed into how to maximize the rigor and reproducibility of integrated human-machine citizen science projects, as well as to address challenges in crowdsourced data provenance, attribution, bias, integrity, and verification. For example, citizen science combined with augmented reality can improve data quality and quantity by helping to accurately target optimal data collection locations, including natural environment landmarks or locating animals from drone imagery.

**Privacy, Security, and Trust.** As the government begins to rely on citizen science data to inform decision making, agencies will need to identify, track, and mitigate possible cybersecurity and "social cybersecurity" threats in these systems. Research is needed to characterize and develop strategies to combat computer-mediated manipulation of citizen science volunteer behavior, as well as the potential spread and impact of dis-/misinformation and false data through citizen science platforms, mobile apps, and observations (video, text, speech). In addition, research is needed into the ethical, legal, and social implications (ELSI) of combining citizen science and AI systems. New ethical frameworks, processes, and updated Institutional Review Board (IRB)/Human Subjects Research (HSR) protocols are likely to be needed. For instance, data collected by or about citizen science volunteers may create privacy and ethical challenges, as well as obligations under the 1974 Privacy Act and HIPPA, among other statutes.

**Citizen Science Data Cyberinfrastructure and the Cloud.** Open science is research that is collaborative, transparent, and reproducible and where the outputs, including data, software, and publications, are made publicly available. To align citizen science and crowdsourcing with Open Science and Open Data objectives, Federal Agencies will need to integrate citizen science into their cloud strategies. Citizen science data are scattered in small ponds across projects. Providing and maintaining cyberinfrastructure and cloud computing for citizen science data, along with ontological, exchange, and metadata standards, could increase open data access and use for distributed volunteers, scientific organizations, and the public, as well as ensure the long-term sustainability and scalability of projects.

**Broadening Participation.** Participation in citizen science is currently skewed demographically and geographically, reflecting different motivations and capacity to participate. For example, data from rural and underrepresented communities are often not present in distributed observational projects. Broader engagement of the public through citizen science and crowdsourcing, particularly early in the research-design phase, can help to reduce bias in data and training annotations for AI, enable public shaping and trust in AI, and foster lifelong interest in science. Research is needed to create adaptive computing solutions that can help remove barriers to participation in citizen science, enabling and equitably incentivizing participatory science, providing access across age, geographic locations, cultures, abilities, diversity of backgrounds, and varieties of motivations. In addition, research is needed on systems enabled by AI, open data practices, and data-visualization tools, to support the public in making use of the results of their observations and thereby taking a seat at the table in setting the agenda for research and democratizing science[7].

**Training, Education, and Learning.** A commonly-reported motivation to participate in citizen science is the opportunity to learn about science. Research is needed on the use of AI with citizen science to improve training, education, and learning. For example, AI-techniques could promote learning by answering common volunteer questions or recommending additional resources. AI-initiated training modules could help to improve volunteer data quality. Studies are also needed to better understand how to foster the synergistic learning that can arise from human-AI teaming when both partners respond adaptively to each other to improve their team performance.

---

[7] NASEM. *Learning Through Citizen Science: Enhancing Opportunities by Design*. (2018) National Academies Press. DOI 10.17226/25183

**Programmatic Recommendations**

This research agenda will require a combination of computer science, data science, social science, and transdisciplinary research, as well as federal agency investments in citizen science and AI across a broad range of federal grant and procurement programs. Other programmatic recommendations include:

- **Interagency Coordination**. To advance citizen science and AI in transdisciplinary research, and to improve interagency coordination and cross pollination of research ideas and best practices, support for the FedCCS and *Citizenscience.gov* should be moved from GSA TTE to the GSA Office of Government-wide Policy. In addition, close coordination between the FedCCS, GSA, the White House (OSTP), Office of Management and Budget (OMB), and the Networking & Information Technology R&D (NITRD) program on citizen science and AI R&D will be essential to success.

- **Regulatory Issues**. The Paperwork Reduction Act (PRA) requires that federal agencies seek formal approval from OMB before collecting information from the public. This approval process has taken upwards of twelve months for many federal citizen-science projects, inhibiting innovation. The PRA was not designed for new computational capabilities that enable federal agencies to collect ongoing scientific data streams from millions of volunteers. We recommend that OMB find ways to adapt its procedures to these new collection techniques. There also is an urgent need to update the 2010 OMB memo that facilitated scientific research by streamlining the PRA information clearance process to include crowdsourced and citizen science data collections[8].

- **Public-Private Partnerships**. Federal citizen science initiatives would benefit from developing public-private partnerships with academia, industry, non-profits, and professional associations to advance citizen science and AI research, and to develop the "connective tissue" necessary to link projects with similar research questions to address national and global challenges.

- **Evaluation and Competes Act Reporting.** Federal agencies should include citizen science and AI as a budget line item, and inventory and evaluate the performance of not only federally managed citizen-science projects, but also of federally funded citizen science research projects to assess and better respond to emerging research trends in the field.

We have an opportunity to harness the curiosity and creativity of the American people to address some of the world's most pressing challenges, such as climate change, the UN Sustainable Development Goals, biodiversity, water and food security, and pandemics. **The grand challenge will be coordinating across citizen science and AI research projects with shared interests on national and global scales.** To achieve this challenge, we need to lead at the federal level with cross-cutting strategies, supportive policies, and a research agenda that will foster unprecedented capabilities to tackle these challenges, not as a government or as individuals, but as a collective national resource of human ingenuity.

*This white paper is part of a series of papers compiled every four years by the CCC Council and members of the computing research community to inform policymakers, community members and the public on important research opportunities in areas of national priority. The topics chosen represent areas of pressing national need spanning various subdisciplines of the computing research field. The white papers*

---

[8] Gellman, R. *Crowdsourcing, Citizen Science, and the Law*. (2016) The Wilson Center. Available at: https://www.wilsoncenter.org/sites/default/files/media/documents/publication/executive_summary_gellman.pdf

*attempt to portray a comprehensive picture of the computing research field detailing potential research directions, challenges and recommendations.*

*This material is based upon work supported by the National Science Foundation under Grant No. 1734706. Any opinions, findings, and conclusions or recommendations expressed in this material are those of the authors and do not necessarily reflect the views of the National Science Foundation.*

*For citation use: Shanley, L.A., Fortson, L., Berger-Wolf, T., Crowston, K., and Michelucci, M. (2021) Imagine All the People: Citizen Science, Artificial Intelligence, and Computational Research. Washington: D.C.: Computing Community Consortium (CCC).*

**ADDENDUM: Citizen Science + AI/ML Projects Align with Biden-Harris Administration Priorities**

**Health:** Projects like *[StallCatchers,](#)* *[EternaBrain](#)*, and *[MoyoHealth](#)* are combining citizen science and AI to advance diagnostic research, enable patient-led research, and improve healthcare delivery. For example, Stall Catchers crowdsources the analysis of Cornell University's Alzheimer's research data ten times faster and with better data quality than laboratory technicians. Over 30,000 online volunteers have analyzed more than 2.5 million blood vessels flagged by AI algorithms, enabling the discovery of a compound that will be studied in human clinical trials as a potential first-ever treatment for alleviating reduced brain blood flow in Alzheimer's patients. Combining community science with agile development, researchers designed and evaluated an open source infrastructure that integrates electronic health records with data collected with a mobile app called *[MoyoHealth](#)*. Co-created with young adults in African American communities in the South, a group at high risk for heart disease, the Moyo app allows users to track risk factors, such as high blood pressure, low physical activity, poor diet, and obesity. By integrating data from phone sensors, wearables, and external databases on a cloud platform, the app allows users (and their doctors) to track information about their health habits alongside environmental factors, such as weather and pollution, from NASA and NOAA Earth observation data.

**COVID-19:** Numerous citizen science and crowdsourcing [projects](#) are addressing the SARS-cov-2 virus. The *[COVID Symptom Study](#), for instance, uses a mobile app to* engage more than 4.5 million people in the US, UK, and Sweden to track COVID-related symptoms, adherence to vaccinations, and other related data. This project thus far has signed up 859,000 people to participate in vaccine clinical trials and resulted in 20 peer-reviewed scientific papers. It was the [first to identify loss of smell and taste](#) as common symptoms of COVID-19. Using data from the mobile app, researchers developed an [AI model](#) that is able to predict whether a person has the COVID-19 infection without the need for testing.

**Racial Equity:** In projects like *[Antislavery Manuscripts](#)*, *[Mapping Prejudice](#)*, *[Scotus Notes,](#)* and *[Douglass Day](#),* the volunteer crowdsourced transcription of historical handwritten documents, such as property deeds, court records, and even letters between abolitionists, have enabled researchers to trace the systemic nature of racism in this country, and informed our understanding of ways to address racial inequities. With advances in Natural Language Processing, many of these projects are incorporating AI to generate suggested transcriptions which are then collaboratively edited by citizen historians.

**Climate Change and Conservation:** Through projects like *[iNaturalist](#)*, *[iDigBio](#)*, *[Snapshot Safari](#)*, and *[Wildbook](#)*, citizen scientists can help track, quantify, and understand the impacts of climate change by observing and recording seasonal changes in plants and animals, from frog calls to flower blooms. For example, *[Nature's Notebook](#)*, a project of the USA National Phenology Network, has more than 15,000 volunteers following carefully developed guidelines to track life stages of nearly 1500 species of plants and animals over the long term. AI is increasingly being used for image and audio analysis in biodiversity monitoring, including incorporation of citizen science images and recordings, as well as feedback for citizen scientists.

**Sustainable Development:** Citizen science and crowdsourcing are helping to address the UN Sustainable Development Goals (SDGs) [in many ways](#). The *[Humanitarian OpenStreetMap Team](#)* (HOTOSM), for example, has engaged more than 300,000 volunteers to assist humanitarian and disaster relief efforts through satellite image analysis and AI, mapping more than 101,958,054 buildings and 2,392,572 roads. HOT's work contributes to the achievement of the SDGs, including disaster risk reduction, clean energy, water and sanitation, and more.